\documentclass[12pt]{iopart}

\pdfoutput=1

\usepackage{iopams}
\usepackage{dcolumn}
\usepackage{color}
\usepackage[final]{graphicx}
\usepackage{bm}
\usepackage{comment}

\newcommand{\be}{\begin{equation}}
\newcommand{\ee}{\end{equation}}
\newcommand{\bea}{\begin{eqnarray}}
\newcommand{\eea}{\end{eqnarray}}
\newcommand{\bit}{\begin{itemize}}
\newcommand{\eit}{\end{itemize}}

\newcommand{\bfl}{\begin{flushright}}
\newcommand{\efl}{\end{flushright}}

\newcommand{\non}{\nonumber \\}
\newcommand{\nonu}{\nonumber}

\newcommand{\ra}{\rangle}
\newcommand{\la}{\langle}

\begin{document}

\title{Shapiro effect in atomchip-based bosonic Josephson junctions}

\author{Julian Grond$^{1,2,3,4}$, Thomas Betz$^1$, Ulrich Hohenester$^3$, Norbert J. Mauser$^2$, J\"org Schmiedmayer$^1$,
and Thorsten Schumm$^{1,2}$}
\address{$^1$ Vienna Center for Quantum Science and Technology, Atominstitut, TU Wien, 1020 Vienna, Austria}
\address{$^2$ Wolfgang Pauli Institut c/o Fak. Mathematik, Universit\"at Wien, Nordbergstrasse 15, 1090 Vienna, Austria}
\address{$^3$ Institut f\"ur Physik,
  Karl--Franzens--Universit\"at Graz, Universit\"atsplatz 5,
  8010 Graz, Austria}
\address{$^4$ Theoretische Chemie, Physikalisch--Chemisches Institut, Universit\"at Heidelberg,
Im Neuenheimer Feld 229, 69120 Heidelberg, Germany}

\begin{abstract}
We analyze the emergence of Shapiro resonances in tunnel-coupled Bose-Einstein condensates, realizing a bosonic Josephson junction. Our analysis is based on an experimentally relevant implementation using magnetic double well potentials on an atomchip. In this configuration the potential bias (implementing the junction voltage) and the potential barrier (realizing the Josephson link) are intrinsically coupled. We show that the dynamically driven system exhibits significantly \emph{enhanced Shapiro resonances} which will facilitate experimental observation. To describe the systems response to the dynamic drive we compare a single-mode Gross-Pitaevskii (GP) description, an improved two-mode (TM) model and the self-consistent multi-configurational time dependent Hartree for Bosons (MCTDHB) method. We show that in the case of significant atom-atom interactions or strong driving, the spatial dynamics of the involved modes have to be taken into account, and only the MCTDHB method allows reliable predictions.
\end{abstract}






\maketitle

\section{Introduction \label{sec:intro}}

Coherent tunnelling dynamics of macroscopic many-body quantum states through a classically forbidden barrier is one of the most striking manifestations of quantum physics. In the presence of an external drive, such systems exhibit the \emph{Shapiro-effect}, also known as \emph{photon-assisted tunneling}~\cite{barone:82,eckardt:05}. Shapiro resonances have been studied extensively in the context of solid-state superconducting Josephson junctions where a resonant modulation of the energy bias leads to a DC current across the junction \cite{tinkham:97}. Such resonances can be used to exactly quantify an applied DC potential difference across the junction and are nowadays used to implement a voltage standard. 

A similar effect occurs in tunnel-coupled Bose-Einstein condensates (BEC) in double well potentials, where a resonant modulation of the bias energy lets the condensates oscillate between both wells~\cite{eckardt:05}. Such dynamics could be used for an accurate measurement of the chemical potential difference across the tunnel barrier~\cite{kohler:03}. Despite the close analogy to the superconducting system, the dynamics of the \emph{bosonic Josephson junction} are strongly influenced by atom-atom interactions~\cite{Raghavan1999}. Additionally the presence of a finite atom reservoir inhibits the observation of a true DC atom flux.

In recent years much effort has been devoted to the theoretical and experimental study of the tunnelling dynamics of driven Bose-Einstein condensates~\cite{morsch:06}. In optical lattice experiments, arrays of bosonic Josephson junctions have been realized \cite{cataliotti:01}. Shaking of the lattice allowed to control the effective tunnel coupling~\cite{lignier:07} and the superfluid to Mott-insulator transition~\cite{zenesini:09}. This effect was also used for dynamical localization~\cite{eckardt:09}. The direct analogy to the superconducting Josephson junction has been realized in an all-optical double well potential. Here the influence of atom-atom interactions has been evidenced, leading to new dynamical modes~\cite{albiez:05} and sub-poissonian number statistics~\cite{esteve:08}.

In this work we analyze the Shapiro effect for magnetic double well potentials on atomchips in view of a recent experimental realization of a bosonic Josephson junction~\cite{Betz2011}. We focus on the experimentally relevant situation where a modulation of the energy bias between the left and right BEC is accompanied by a concurrent modulation of the tunnel coupling. As we show, this leads to a significant enhancement of the Shapiro resonances compared to the conventional driving~\cite{eckardt:05}, which will facilitate experimental observation.

Shapiro resonances have been theoretically analyzed in~\cite{grifoni:98,eckardt:05} within a two-mode (TM) model employing static condensate wave functions~\cite{milburn:97,javanainen:99}. For the configuration investigated in this work it is important to take into account the spatial dynamics, as transverse excitations might occur due to the external drive, which have a strong influence on the tunneling dynamics. We therefore employ the \emph{multi-configurational time dependent Hartree equations for Bosons} (MCTDHB) method~\cite{alon:08,meyer:09}, which is a self-consistent, essentially exact framework, taking into account the full dynamics of the system. In previous work, the MCTDHB method applied to bosonic Josephson junctions allowed to discover dynamics very different to that predicted by the Gross-Pitaevskii equation and the two-mode model, such as an \emph{inverse} splitting regime~\cite{streltsov:07}, the decay of macroscopic quantum self-trapping~\cite{sakmann:09}, and violation of time-reversal symmetry when switching interactions from repulsive to attractive in bosonic Josephson junctions~\cite{sakmann.pra:10}. In this work we study another case, where for typical interaction strengths the Shapiro resonances predicted from MCTDHB appreciably differ from the simpler models.

The paper is organized as follows: We start with the physical system and its dynamic description in section~\ref{sec:system}. In section~\ref{subsec:dw} we introduce the realization of the double well potential and how we implement the driving, and in section~\ref{subsec:model} the three models which will be used and compared in this work: the Gross-Pitaevski (GP) equation, the two-mode model, and MCTDHB. In section~\ref{sec:ref} the non-interacting driven system is discussed, and results which demonstrate \emph{enhanced Shapiro resonances} are shown. It includes also a short discussion of how to choose experimental parameters on order to observe the effect. Results for the interacting case are presented in section~\ref{sec:interacting}. First we show how the resonances shift due to interactions in section~\ref{subsec:shifts}. A discussion of the different dynamical models, as well as the damping of the number imbalance is finally discussed in section~\ref{subsec:damp}.

\section{Physical system\label{sec:system}}

\subsection{Double well potential \label{subsec:dw}}

The following analysis is motivated by a recent experimental realization of a bosonic Josephson junction on an atomchip~\cite{Betz2011}. A symmetric double well potential is generated by a combination of static and oscillating magnetic near fields, making use of dressed adiabatic states~\cite{schumm:05,lesanovsky:06,hofferberth:07b} \footnote{The considerations made in this work are equally valid for double well systems based on solely static magnetic fields or on optical dipole potentials.}. The system consists of two elongated traps with strong atomic confinement ($\omega_{\perp}\sim2\pi \cdot 2$ kHz) in the $x$ and $y$ (transverse) directions, and a very weak confinement along the $z$ (longitudinal) direction with $\omega_{x,z}/\omega_{z}\sim100$. Tunnelling dynamics takes place along the coupling direction $x$ which connects the two potential minima of the double well. The system is assumed to remain in the (many-particle) ground state in the two orthogonal directions $y$ and $z$ which do not contribute to the dynamics. The distance $d$ between the double well minima and the height of the tunnel barrier are adjusted by controlling the amplitude of the oscillating field component which is in the radio frequency (RF) range. This RF amplitude represents the dynamical control parameter $\lambda(t)$ used to drive the Shapiro resonances.

To implement a potential difference ("voltage") between both wells, the double well can be tilted in space (making use of gravity) \cite{schumm:05,Baum2010}. Further possibilities are a spatial inhomogeneity in the amplitude of the RF field or employing additional forces like DC electric near fields \cite{krueger:03} or local optical dipole potentials \cite{albiez:05,leblanc:11}. The resulting potential is given by equation~10 of~\cite{lesanovsky:06} and can be approximately described by a symmetric fourth order polynom $V_{\lambda}(x)$ for the double well with an additional linear gradient $g$ that implements the tilt: 
\bea
V_{\lambda,g}(x)&=&V_{\lambda}(x)+g\cdot x\non
&=&c_1(\lambda)\left[x-d(\lambda)/2\right]^2+c_2(\lambda)\left[x-d(\lambda)/2\right]^4+g\cdot x \,.
\eea
Note that the control parameter $\lambda$ (RF amplitude) affects both the separation of the double well minima ($d(\lambda)$) and the steepness of the confinement (expressed by the coefficients $c_1d(\lambda)$ and $c_2d(\lambda)$). By construction, a larger separation coincides with a higher potential barrier, both reducing coupling, making $\lambda$ the most sensitive parameter in the system.

\begin{figure}
\centering
 \includegraphics[width=.8\columnwidth]{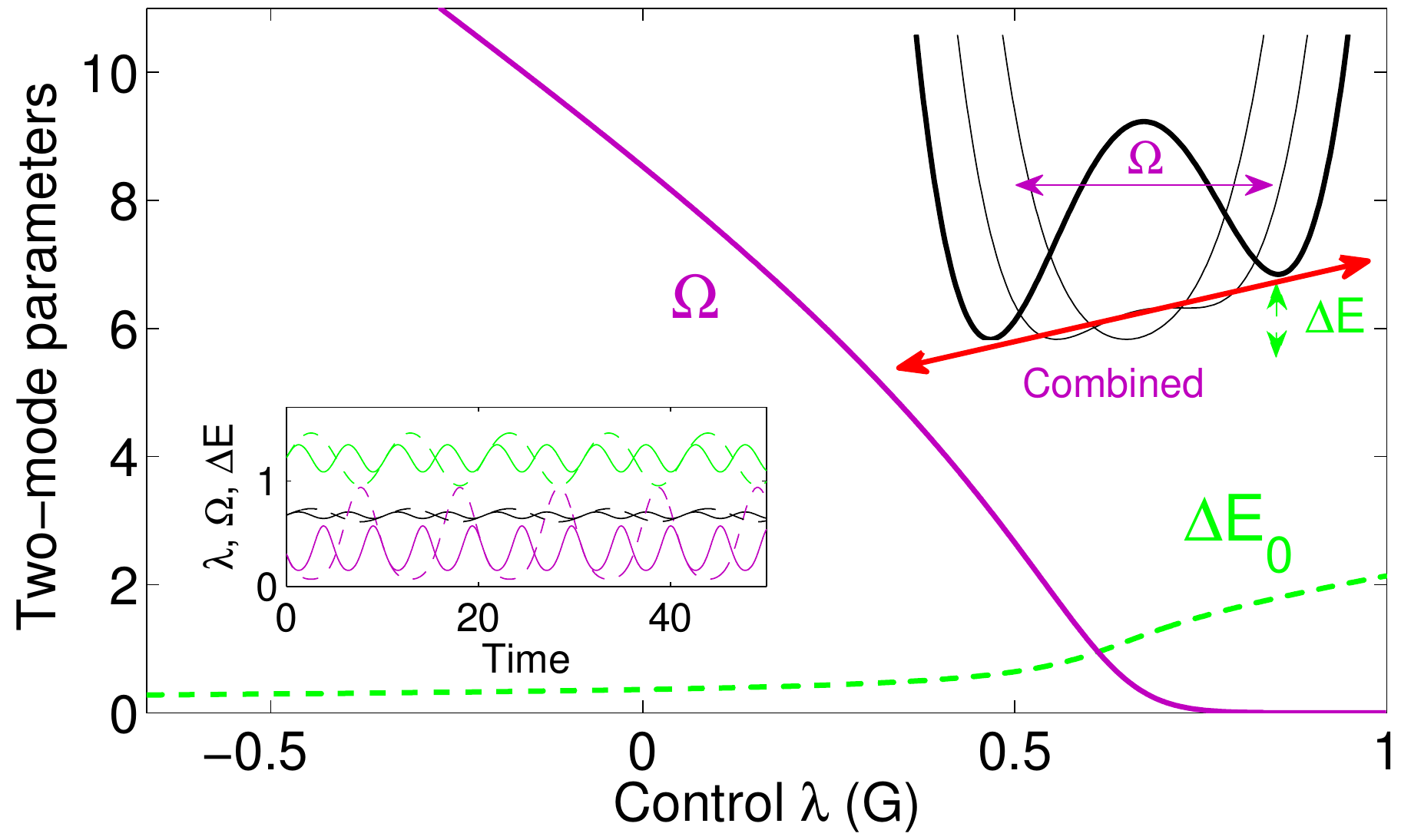}
\caption{ $\Omega$ and $\Delta E$ versus $\lambda$, calculated using eigenstates of the Schr\"odinger equation. Upper right inset: Driving geometry. Lower left inset: When the splitting parameter $\lambda$ (black lines) is driven periodically, both $\Omega$ (violet lines) and $\Delta E$ (green lines) oscillate. Solid line: first resonance $n=1$ for the non-interacting system, with parameters as in text. Dashed line: second resonance $n=2$ (with larger amplitude). \label{fig:static}}
\end{figure}

The most straightforward implementation of a Shapiro experiment would be to dynamically modulate the potential gradient $g$. However in a purely magnetic implementation of the double well as in~\cite{Betz2011} this would also result in a significant spatial displacement of the trap minima, which might lead to excitations and uncontrolled dynamics. We therefore focus the analysis on the effects of a periodic modulation of the RF amplitude (splitting parameter $\lambda(t)$) keeping $g$ fixed. This results in double well minima moving along a potential slope as indicated in figure~\ref{fig:static} (upper right inset). As we will show, modulating both the tunnel coupling and the potential difference leads to a significant enhancement of the observed Shapiro resonances.

\subsection{Model \label{subsec:model}}

\subsubsection{Many-body Hamiltonian}

In this work we consider the dynamics in the splitting direction $x$, and neglect any dynamics in the other directions $y$ and $z$~\cite{grond.pra:09b}. The dynamics of the atoms in the double well is then governed by the many-body Hamiltonian~\cite{dalfovo:99,leggett:01}
\begin{equation}\label{eq:ham}
  \hat H(t)=\int\left[\hat\Psi^\dagger(x)\hat h^{\lambda}(x)
  \hat\Psi(x)+\frac{U_0}{2}\hat\Psi^\dagger(x)\hat\Psi^\dagger(x)\hat\Psi(x)\hat\Psi(x)
  \right]\,dx\,,
\end{equation}
where $\hat\Psi(x)$ is the bosonic field operator, and $\hat h^{\lambda}(x)=-\frac{1}{2}\nabla^2+V_{\lambda,g}(x)$ is the \emph{bare} Hamiltonian.  $U_0$ is the effective 1D interaction strength, obtained by integrating out the other spatial directions for the ground state~\cite{grond.pra:09b}. The 1D approximation is reasonable since typically the dynamics in the other directions decouples to a good extent whenever the 3D-Potential factorizes, i.e.,  $V(\mathbf{r})=V_x(x)+V_y(y)+V_z(z)$.

Direct solution of the dynamics due to this Hamiltonian is not possible for more than a few atoms, and thus we have to use approximation schemes. In this paper we will employ and compare several types of approximations which we will outline in the following. They can be considered as different ways of restricting the field operator $\hat\Psi(x)$ to a small number of modes. 

\subsubsection{Gross-Pitaevskii (GP) equation}

The mean field dynamics~\cite{dalfovo:99} is obtained by restricting the field operator to a single mode function, $\hat\Psi(x)=\hat a_0\phi_0(x)$. The Heisenberg equation of motion for $\hat\Psi(x)$, obtained from equation~\ref{eq:ham}, yields the GP equation 
\be
i\dot{\phi}_0(x,t)=\Bigl(\hat h^{\lambda}(x)+U_0 (N-1)|\phi_0(x,t)|^2\Bigr)\phi_0(x,t)\;.
\ee
For a BEC in a double well, the GP wave function can be written as $\phi_0(x)=[\phi_L(x)+\phi_R(x)]/\sqrt{2}$, where $\phi_L(x)$ ($\phi_R(x)$) is localized in the left (right) well, and thus the many-body wave function reads
\be
\Psi(x)=\Bigl([\phi_L(x)+\phi_R(x)]/\sqrt{2}\Bigr)^N=\frac{1}{2^{N/2}}\sum_{k=0}^N\Biggl( \begin{array}{c} N\\ n\end{array} \Biggr)\phi_L^n(x) \phi_R^{N-n}(x)\,.
\ee
The GP equation assumes that the number distribution between the left and right localized condensate has binomial number fluctuations $\sqrt{N}$ (perfectly ''coherent state``~\cite{pitaevskii:01}).
Since for the initial state we consider in this work always a condensate localized in one well, the GP equation will well describe the short-time-dynamics. At longer times, however, a single-mode description will not be valid anymore. 

\subsubsection{Two-mode (TM) model}

In  a double well it is more natural to use a basis which comprises two modes instead of one: $\hat \Psi^{\lambda}(x)=\hat a_L\phi_L^{\lambda}(x) + \hat a_R\phi_R^{\lambda}(x)$. Here, the localized mode functions $\phi_L^{\lambda}(x)$ and $\phi_R^{\lambda}(x)$ are depend on the control parameter $\lambda$. $\hat  a_L$ ($\hat a_L^{\dagger}$) annihilates (creates) an atom in the left well, and similar for $\hat a_R$ ($\hat a_R^{\dagger}$). 

There exist several schemes for the choice of the modes $\phi_L^{\lambda}(x)$ and $\phi_R^{\lambda}(x)$. The simplest one uses superpositions of the two lowest lying eigenstates $\phi_g^{\lambda}(x)$ and $\phi_e^{\lambda}(x)$ of the single-particle Schr\"odinger equation of the symmetric potential $V_{\lambda}(x)$, which have gerade and ungerade symmetry: $\phi_{L,R}^{\lambda}(x)=[\phi_g^{\lambda}(x)\pm\phi_e^{\lambda}(x)]/\sqrt{2}$~\cite{milburn:97,javanainen:99}. The Hamiltonian reads in terms of pseudo-spin operators:
\be\label{eq:hamTM}
H=-\Omega(t)\hat{J}_x+\Delta E(t)\hat{J}_z+2\kappa(t)\hat{J}_z^2\;.
\ee
Hereby, $\hat{J}_z=\frac 12(\hat a_L^\dagger \hat a_L^{\phantom\dagger}-\hat a_R^\dagger \hat a_R^{\phantom\dagger})$ measures the atom number difference between left and right well, and $\hat J_x=\frac 12(\hat a_L^\dagger \hat a_R^{\phantom\dagger}+ \hat a_R^\dagger \hat a_L^{\phantom\dagger})$ promotes an atom from the left to the right well and vice versa. Equation~\ref{eq:hamTM} depends on the generic parameters $\Omega(t)$, $\Delta E(t)$ and $\kappa(t)$, which denote the tunnel coupling, energy bias, and nonlinear interaction energy, respectively. They are given as 
\bea
&&\Omega(t)=-\int dx \phi_L^{*,\lambda}(x)\hat h^{\lambda}(x)\phi_R^{\lambda}(x) + h.c.\,,\\
&&\Delta E(t)=\int dx \phi_L^{*,\lambda}(x)\hat h^{\lambda}(x)\phi_L^{\lambda}(x)-\int dx \phi_R^{*,\lambda}(x)\hat h^{\lambda}(x)\phi_R^{\lambda}(x)\,,\non
&&\kappa(t)=\frac{U_0}{2}\int dx|\phi_{L,R}^{\lambda}(x)|^4\,.\nonu
\eea
The time-dependence in these parameters is due to a time-dependent control $\lambda(t)$. Tunnel coupling and energy bias between left and right condensates are shown in figure~\ref{fig:static} over a wide range of $\lambda$ from the unsplit to the split case. \footnote{We use units where $\hbar=1$, mass of a $^{87}$Rb atom $m=1$, and atom energy and time is scaled by the confinement length $a_{ho} = \sqrt{\hbar/(m\omega_{ho})}=1$ $\mu m$ and energy $\hbar \omega_{ho}$ of a harmonic oscillator. It follows that the units of time and energy are, respectively, $1/\hbar\omega_{ho}=1.37$ ms and $\hbar \omega_{ho}=2\pi\cdot 116.26$ Hz.} While the tunnel coupling decreases with increased splitting, the energy bias becomes larger. Typically, we choose  a constant $\kappa\approx U_0/2$, which is a good approximation because this value depends less crucial on the shape of the modes as compared to $\Omega(t)$ and $\Delta E(t)$. 

An \emph{improved} two-mode model~\cite{ananikian:06} can be obtained by using the first and second self-consistent states of the GP equation $\phi_g^{\mathrm{GP},\lambda}(x)$ and $\phi_e^{\mathrm GP,\lambda}(x)$ with self-consistent energies $E_g^{\mathrm GP,\lambda}$ and $E_e^{\mathrm GP,\lambda}$, respectively. The tunnel coupling of the improved model $\Omega^{(I)}$ is given by the energy difference $E_e^{\mathrm GP,\lambda}-E_g^{\mathrm GP,\lambda}$ and a shift due to interactions. \footnote{Other, more subtle corrections to the standard two-mode model (e.g., to $\kappa$), as suggested by Bergeman \emph{et al.}~\cite{ananikian:06}, will not be used here since it does not lead to significant improvements.} The parameters for the improved model are then given by
\bea\label{eq:improved}
&&\Omega^{(I)}(t)=E_e^{\mathrm GP,\lambda}-E_g^{\mathrm GP,\lambda}-U_0 N/2\Biggl(\int dx|\phi_{e}^{\mathrm GP,\lambda}|^4-\int dx|\phi_{g}^{\mathrm GP,\lambda}|^4\Biggr)\,,\\
&&\Delta E^{(I)}(t)=\int dx [\phi_L^{\mathrm GP,\lambda}]^*\hat h^{\lambda}\phi_L^{\mathrm GP,\lambda}-\int dx [\phi_R^{\mathrm GP,\lambda}]^*\hat h^{\lambda}\phi_R^{\mathrm GP,\lambda}\,.
\eea
In the following we always use the improved two-mode model, unless stated otherwise.

\subsubsection{Two-mode model with self-consistent orbitals and occupations}

A dynamical description including also the spatial dynamics can be obtained by the \emph{multi-configurational time dependent Hartree equations for Bosons} (MCTDHB)~\cite{alon:08}, which represents a framework where the two modes are included self-consistently. Using two time-dependent orbitals has the crucial advantage that transverse spatial excitations can be included, in contrast to a model with two fixed orbitals.

A full discussion of the working equations of MCTDHB~\cite{alon:08,grond.pra:09b} can be found elsewhere. We only sketch here the main ideas of the method. The ansatz for time-dependent modes reads $\hat \Psi^{\lambda}(x)=\hat a_L(t)\phi_L^{\lambda}(x,t) + \hat a_R(t)\phi_R^{\lambda}(x,t)$ for the case of two modes, although in principle the number of orbitals can be chosen at will.\footnote{We simulated MCTDHB with 4 orbitals \cite{MCTDHB,streltsov:10} and found that  2 orbitals allow for a very good description of the physics discussed in this work.} MCTDHB then provides a way to determine the ''best possible`` shape of the orbitals at a given time. This is achieved by formulating an action integral based on the above ansatz, and then using a variational principle. For the orbitals one obtains nonlinear equations, and for the number part two-mode equations similar to the TM model. For the orbitals one obtains nonlinear equations, and for the number part two-mode equations similar to the TM model. Most importantly, the orbital part and the number part are coupled and thus have to be solved self-consistently.

An important quantity is  the one-body reduced density $\la \hat\Psi^{\dagger}(x)\hat\Psi(x)\ra$~\cite{sakmann:08}. It can be diagonalized in terms of natural orbitals $\phi_i(x)$ and natural occupations $\rho_{i}$ ($i=1,2$):
\be
\rho(x)=\sum_{i=1}^2\rho_{i}|\phi_i(x)|^2\;.
\ee
Whenever one natural occupation dominates, we have a BEC~\cite{penrose:56} and one orbital suffices. MCTDHB then coincides with the GP equation. Whenever several eigenstates are finite, we have an $m$-fold fragmented BEC, and $m$-orbitals should be used. Whenever MCTDHB-calculations converge when increasing the number of orbitals, the results can be considered as an exact solution of the many-body Schr\"odinger equation. 

In this work we measure the degree of fragmentation by the difference of the population of the natural orbitals. $(\rho_{11}-\rho_{22})/N=\pm 1$ corresponds to a single BEC, whereas $(\rho_{11}-\rho_{22})/N=0$ corresponds to a fully fragmented system.

\subsection{Observable}

The Shapiro effect in a superconducting Josephson junction is related to a finite DC component of the tunnel current at the resonance frequencies. A similar effect is present in a bosonic Josephson junction~\cite{barone:82}. However, current cannot be measured directly and furthermore the reservoirs, consisting of the atoms in the left or right well, are finite. Therefore, the 'current' changes its sign whenever one reservoir is empty, and in such a manner the atoms oscillate between both wells. 

The initial configuration for the following investigations consists of all atoms in the lower well. We can characterize the Shapiro resonances by a time average of zero atom number difference between the wells. The time averaged imbalance is then, similar as in~\cite{eckardt:05},
\be\label{eq:timeav}
\la J_z\ra_T\equiv\frac{1}{T}\int_0^{T}dt\la J_z\ra(t)\;.
\ee
A value of $0.5$ then corresponds to no population transfer at all, whereas a value close to zero indicates a resonance. 


\section{Shapiro resonances in absence of interactions \label{sec:ref}}

We now discuss the emergence of Shapiro resonances in the atomchip geometry (i.e., by driving the double well separation in the presence of a fixed potential gradient) for the non-interacting case. This system can well be captured within the TM model.

\subsection{Enhanced Shapiro effect \label{subsec:enhanced} }

The time-dependence of the control parameter is chosen as
\be
\lambda(t)=\lambda_0+\lambda_1\sin(\omega t)\,,
\ee
with driving frequency $\omega$. In the following we take $\lambda_0=0.675$, corresponding to a splitting distance of approximately $1\,\mu$m, and a gradient $g=1.5042$, corresponding to $\Delta E_0\approx 2\pi\cdot280$ Hz. Due to the linear relationship between $\lambda$ and the tilt $\Delta E$ at $\lambda_0$ (see figure~\ref{fig:static}), we have to a very good approximation $\Delta E(t)=\Delta E_0+b\cdot \sin(\omega t)$, where $b$ is the driving amplitude of the bias energy. The tunnel coupling instead depends not linearly on the control at $\lambda_0$, but rather in a polynomial fashion. Thus, the general form of the tunnel coupling is 
\be\label{eq:Omm}
\Omega(t)=\Omega_0+\sum_m\Omega_1^{(m)}\sin{(i m\omega t)}\quad (m=1,2,3,...)\,.
\ee 
We now decompose the Hamiltonian, equation~\ref{eq:ham} for $U_0=0$, into $H=H_0(t)+H_1(t)$, with
\bea
H_0(t)=[n\omega+b\sin{(\omega t)}]\hat J_z,\quad H_1(t)=(\Delta E_0-n\omega)\hat J_z-\Omega(t)\hat J_x\,.
\eea
Hereby, $n$ is an integer corresponding to the order of the resonance. Next we transform into an interaction picture \cite{sakurai:94} with respect to $H_0(t)$:
\bea\label{eq:hamI1}
H_1^{(I)}(t)&=&(\Delta E_0-n\omega)\hat J_z -\Omega(t)\Biggl[ e^{ -i[n\omega t+\frac{b}{\omega}\cos{(\omega t)}]} \sum_k|k\ra\la k|\hat J_x|k+ 1\ra\la k+ 1| \non
&&+e^{ i[n\omega t+\frac{b}{\omega}\cos{(\omega t)}]} \sum_k|k\ra\la k|\hat J_x|k- 1\ra\la k- 1|\Biggr]\,,
\eea
where we exploited that $\hat J_x$ couples only neighboring states $|k\ra$ and $|k\pm 1\ra$. Then we insert  the generating function of the ordinary Bessel functions $J_l(z)$
\be
e^{i z \cos{(\omega t)}}=\sum_{l=-\infty}^{\infty}(ie^{i\omega t})^l J_l(z)\,,
\ee
and obtain 
\be\label{eq:hamEff}
H_1^{(I)}=(\Delta E_0-n\omega)\hat J_z-\Omega^{\mathrm{eff}}_n\hat J_{\phi_n}\,,
\ee
with an effective tunnel coupling $\Omega^{\mathrm{eff}}_n$, 
\be\label{eq:Omeff}
\Omega_n^{\mathrm{eff}}=\sqrt{\Omega_0^2 A_n^2+\sum_{m=1}^{\infty}\Omega_1^{(m),2} A_{n-m}^2+2\Omega_0 A_n\sum_{k=0}^{\infty}(-1)^k\Omega_1^{(2k+1)} A_{n-2k-1}}\,,
\ee
where $A_l=J_l\Bigl(\frac{b}{\omega}\Bigr)$. Thus, the contributions to the effective tunnel coupling $\Omega_n^{\mathrm{eff}}$ add up in a vectorial fashion. Further, we have $\hat J_{\phi_n}=\cos{(\phi_n)}\hat J_x+\sin{(\phi_n)}\hat J_y$ where the phase $\phi_n$ has no influence on the considered Shapiro dynamics of $\la J_z\ra(t)$ when initially all atoms are in the lower well. 

We neglected fast oscillating terms and kept only the DC contribution, which is reasonable close to the resonance $\omega\approx\Delta E_0/n$, and whenever the driving frequency $\omega$ is much larger than the coupling $\Omega_0$~\cite{jinasundera:06,grifoni:98}.

Thus, the driven system can be approximately described as an undriven system, but with renormalized tunnel coupling~\cite{tsukada:99,eckardt:05,jinasundera:06}. At the resonance condition $n\omega= \Delta E$, there is no energy bias between the left and the right condensate, and thus the atoms tunnel at rate $\Omega^{\mathrm{eff}}$ (''Shapiro current''). For example, the case $n=0$ corresponds to the trivial resonance of an unbiased system ($\Delta E_0=0$) at $\omega=0$, with Lorentzian line shape characterized by $\Omega^{\mathrm{eff}}$. The effect of the driving is then to modify the depth and the width of the Lorentzian~\cite{tsukada:99}. A system with finite energy bias ($\Delta E_0\neq0$) has similar Lorentzian shaped resonances, but at discrete frequencies $\omega=\Delta E_0/n$.

Most importantly is the oscillatory part of $\Omega(t)$, which gives additional \emph{cross}-contributions to $\Omega^{\mathrm{eff}}$, proportional to $J_{n-m}\Bigl(\frac{b}{\omega}\Bigr)$.  Since $J_{n-1}(x)>J_n(x)$ for small values of $x$, those additional contributions to $\Omega^{\mathrm{eff}}$ can be considerably larger than those due to the first term of equation~\ref{eq:Omeff}, proportional to $J_n\Bigl(\frac{b}{\omega}\Bigr)$, and these contributions can drastically enhance width and depth of the resonances.

\begin{figure}
\centering
 \includegraphics[width=.8\columnwidth]{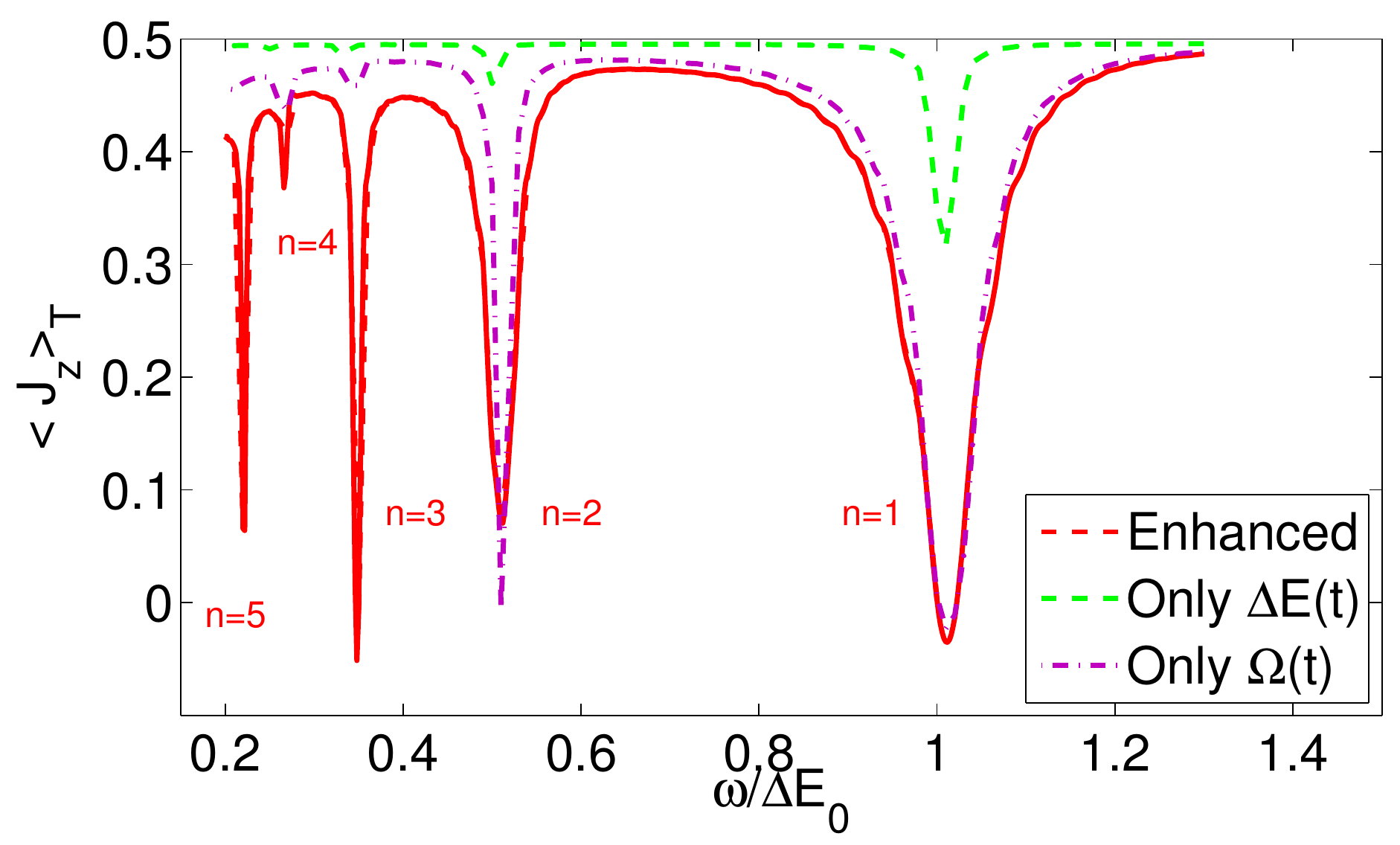}
\caption{Shapiro resonances for the non-interacting case. The time-averaged population imbalance $\la J_z\ra_T$ is plotted, with $T=100$. The driving amplitude $\lambda_1$ depends non-linearly on the driving frequency $\omega$ as $\lambda_1=0.03\cdot\Delta E_0/\omega$. Solid line: \emph{Enhanced Shapiro resonances}.  Dashed line: We compare to the standard Shapiro effect (oscillating $\Delta E(t)$, but constant $\Omega$). Dashed-dotted line: oscillating $\Omega(t)$, but constant $\Delta E$. \label{fig:enhanced}}
\end{figure}

A typical resonance structure is reported in figure~\ref{fig:enhanced} (solid line), showing distinct resonances up to order $n=5$.
We have chosen a driving amplitude  which increases with the order of the resonance as $\lambda_1=0.03\cdot\Delta E_0/\omega$, such that higher resonances become more distinct. The $n=1$ resonance has not only contributions from $J_1\Bigl(\frac{b}{\omega}\Bigr)$, but also from $J_0\Bigl(\frac{b}{\omega}\Bigr)$ since $\Omega_1^{(1)}\neq 0$. The $n=2$ resonance has not only contributions from $J_2\Bigl(\frac{b}{\omega}\Bigr)$, but also from $J_1\Bigl(\frac{b}{\omega}\Bigr)$ since $\Omega_1^{(1)}\neq 0$, and from $J_0\Bigl(\frac{b}{\omega}\Bigr)$ since $\Omega_1^{(2)}\neq 0$. Because $\Omega_1^{(m)}$ is small for $m>2$, the width of the resonances becomes smaller for higher $n$.

In the figure we also compare to the case of an artificially constant $\Omega(t)$ ($\Omega_1^{(m)}=0$ for $m=1,2,3...$) (dashed line), which shows only extremely weak resonances for $n>1$ for the same driving amplitude. This is because the effective tunnel coupling $\Omega^{\mathrm{eff}}$ is proportional to the $n$-th Bessel function $J_n\Bigl(\frac{b}{\omega}\Bigr)$, and our driving amplitude is relatively small as typically $b\sim n/10$. Thus, the contribution of Bessel functions with $n>1$ is very small. 

For the case of an artificially constant $\Delta E(t)$ ($b=0$) (dashed-dotted line), we find very pronounced $n=1$ and $n=2$ resonances.  These resonances are ``trivial`` in the sense that they are due to a cancellation of the oscillatory terms in equation~\ref{eq:hamI1}. The strength of the $n$-th resonance thus corresponds to the magnitude of $\Omega_1^{(m)}$, which is largest for $m=1$ and $m=2$.

Hence, the combined driving with $\Delta E(t)$ and $\Omega(t)$ yields more than the sum of driving with only one of them, and we thus term these resonances \emph{enhanced Shapiro resonances}.

\subsection{Optimal choice of parameters \label{subsec:par}}

We shortly discuss the regime of parameters which are best suited to find clear Shapiro resonances, taking into account limitations posed by an experiment realization.

\subsubsection{Tunnel coupling $\Omega$}
The tunnel coupling $\Omega$, which is determined by the mean double well separation, controls how fast the atoms tunnel. Therefore it should be large enough in order that the averaging period for equation~\ref{eq:timeav} is not too long (typically $\Omega T=15$). 
If $\Omega$ is too large, the resonances are shifted according to $\Delta E\rightarrow\sqrt{\Delta E^2+\Omega^2}$~\cite{eckardt:05}. Due to the larger amplitude Rabi oscillations, the resonances get wider and less clear. Another disadvantage of too large $\Omega$ is that for an interacting system, the initial state will not be localized~\cite{streltsov:06}. This could be compensated by choosing a larger tilt, which has however other disadvantages as discussed below.

\subsubsection{Asymmetry}

A too small asymmetry (linear gradient $g$) is unfavorable in experimental realizations as it leads to tunnelling dynamics already during the system preparation. Furthermore the initial state is not well localized, and no clear resonances can be identified when averaging over time. A too large asymmetry on the contrary requires large driving amplitudes in order to induce a Josephson current.

\subsubsection{Driving amplitude $b$}

We have already seen in section~\ref{sec:ref} that a too small driving amplitude reduces the Shapiro current due to the Bessel function structure. A too large $b$, however, has the effect of broadening the resonances ($\sim J_n(b/\omega)$).

\section{Shapiro resonances in presence of interactions \label{sec:interacting}}

\subsection{Resonance shifts \label{subsec:shifts}}

Shapiro resonances in a many-body system are shifted in frequency due to atom-atom interactions. For weak interactions, as in~\cite{eckardt:05}, the shift is roughly given by the interaction energy $U_0 N$, with the exact resonance frequencies given in terms of elliptic functions~\cite{raghavan:99,eckardt:05}. However, for typical experimental interaction strengths $U_0 N\gtrsim 1$ as we consider here, the resonance frequency depends itself on the driving amplitude $\lambda_1$~\cite{tsukada:99,zhang:08}. 

\begin{figure}
\centering
\begin{tabular}{l l}
 \includegraphics[width=.45\columnwidth]{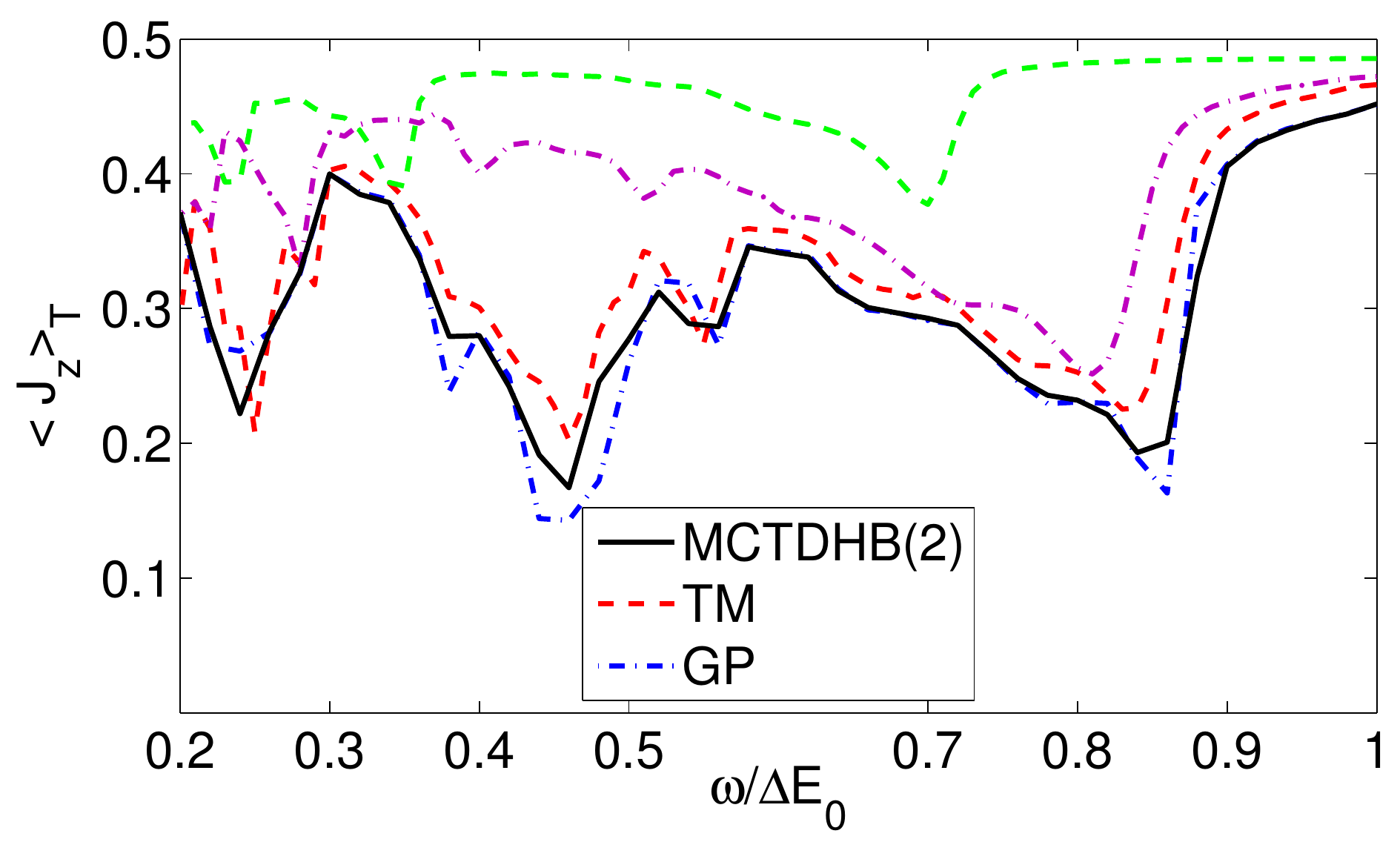}& \includegraphics[width=.45\columnwidth]{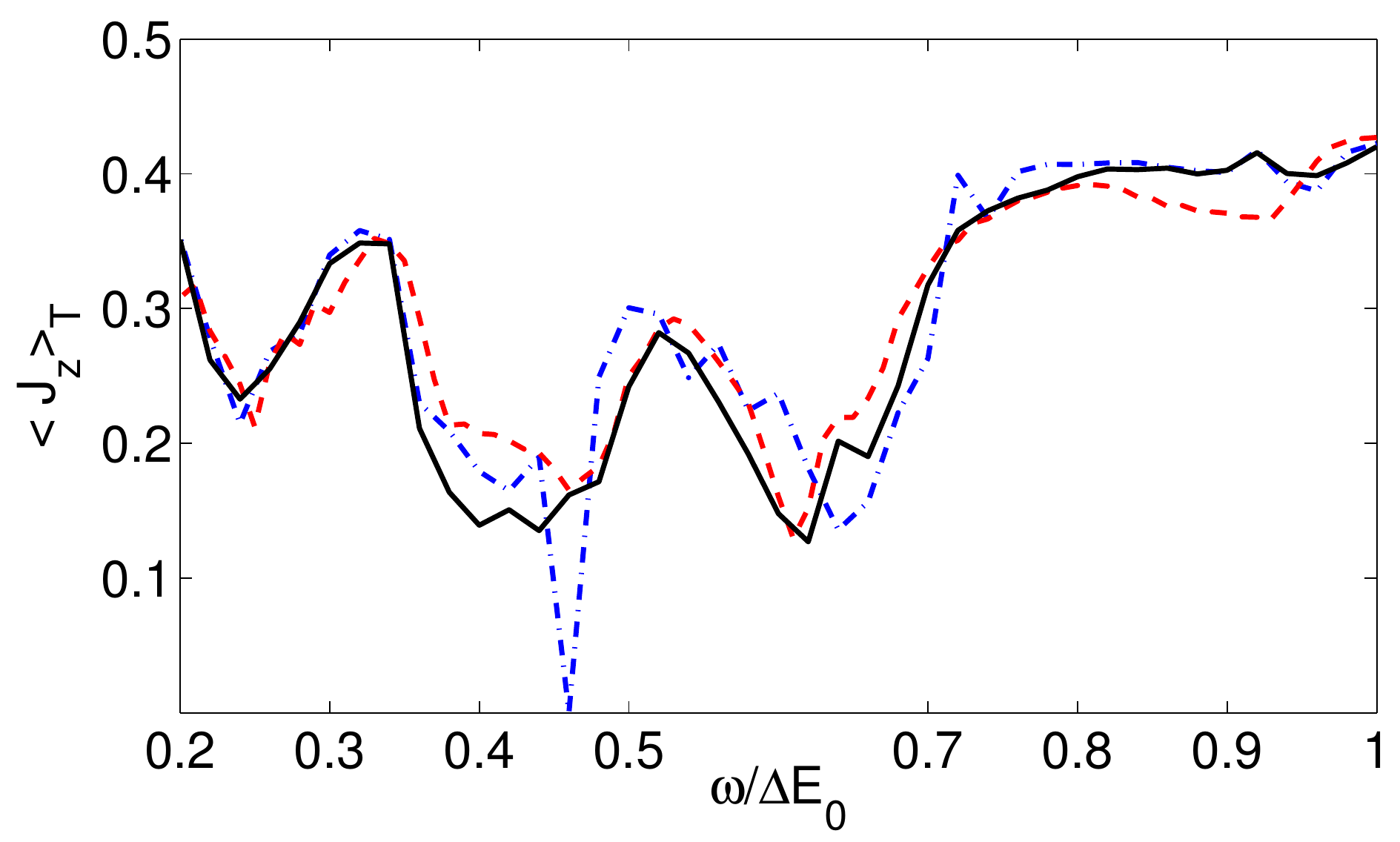}\\
(a)&(b)\\
 \includegraphics[width=.45\columnwidth]{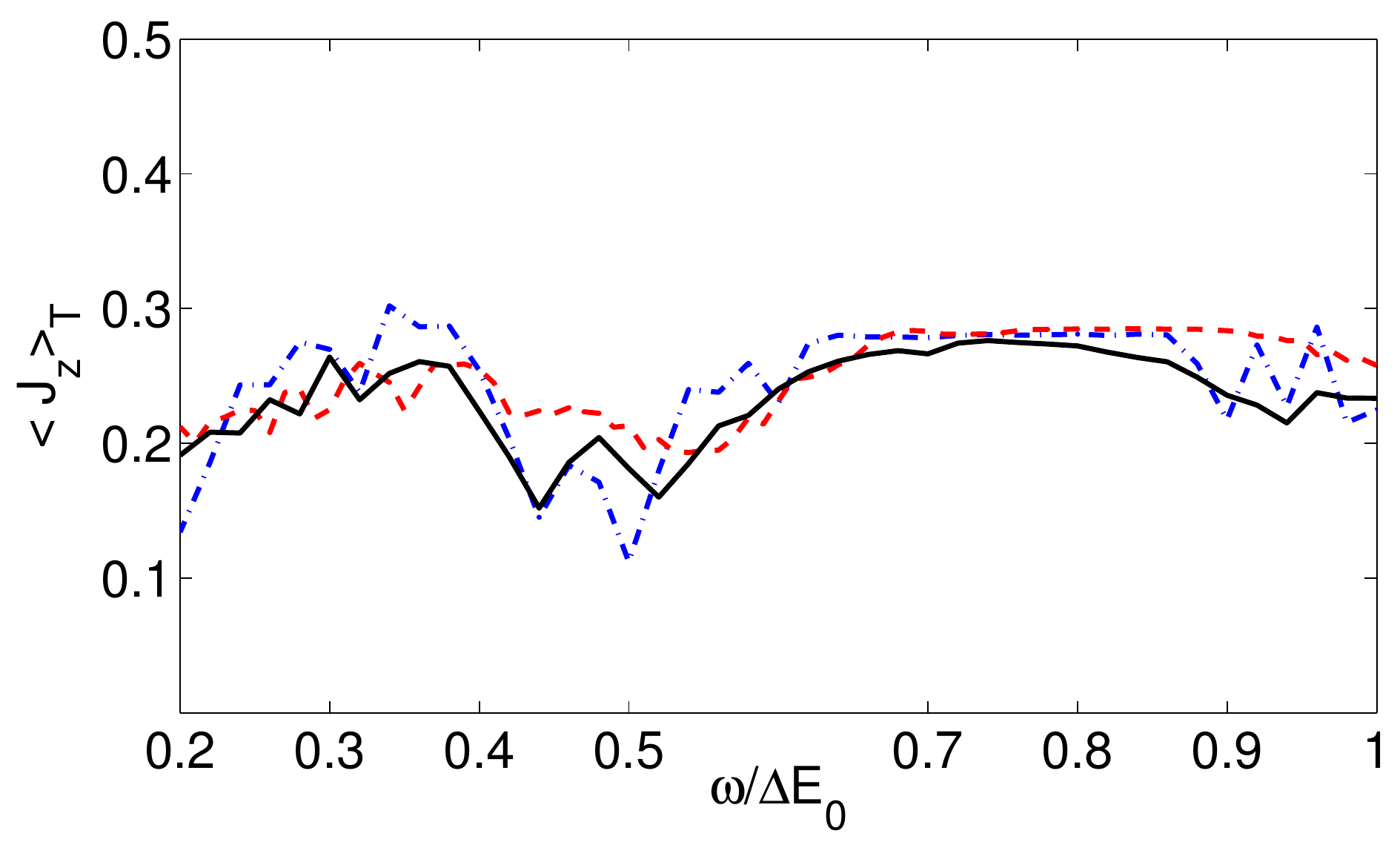}& \includegraphics[width=.45\columnwidth]{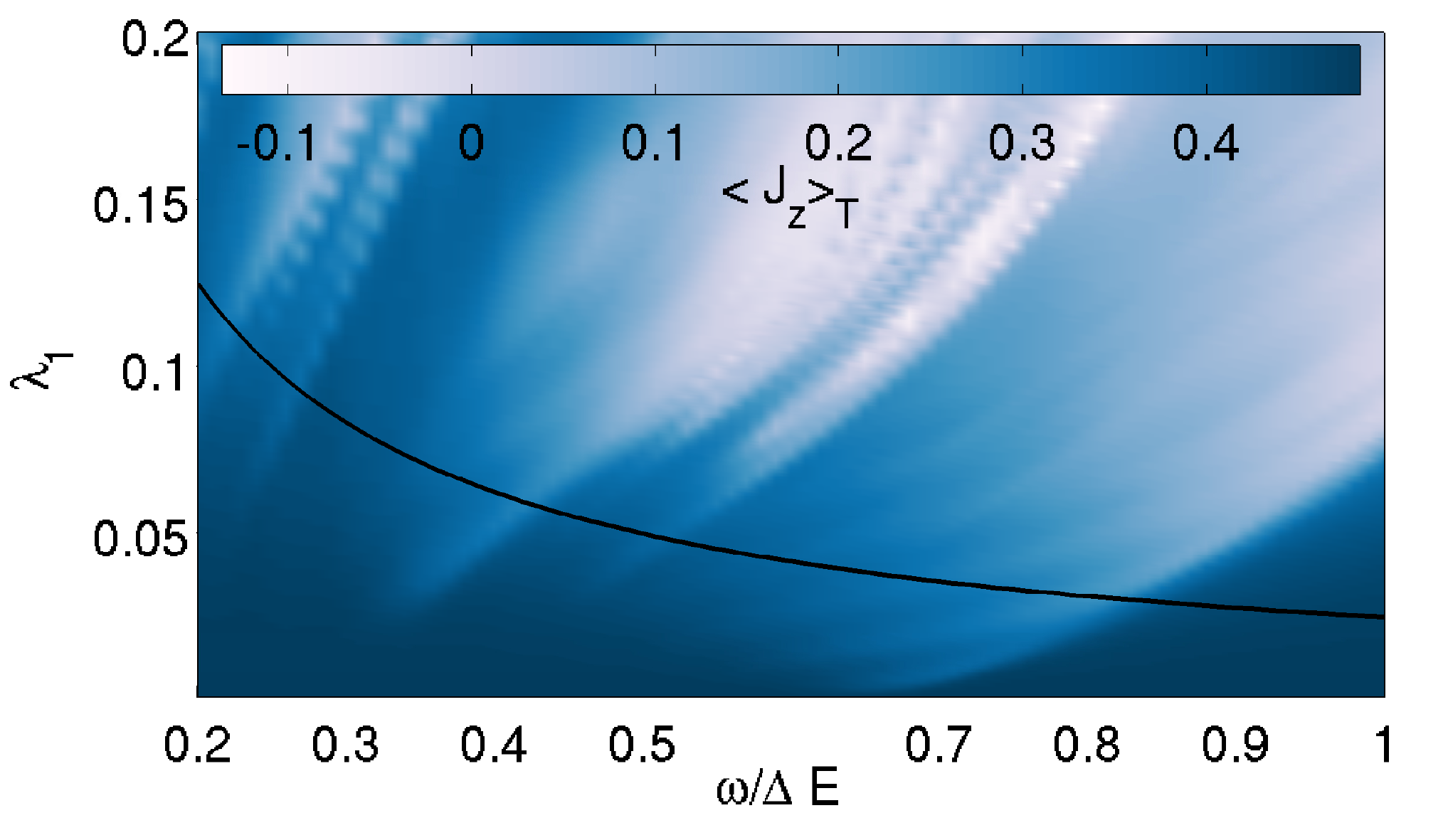}\\
(c)&(d)
\end{tabular}
\caption{Shapiro resonances for interactions (a,d) $U_0 N=1$, (b) $U_0 N=2$, and (c) $U_0 N=4$ ($N=100$). We compare MCTDHB, GP and TM model. (a) We show also the case of constant $\Omega$ or constant $\Delta E$, similar as in figure~\ref{fig:enhanced}. $\lambda_1=0.035\cdot\Delta E_0/\omega$ is used. (b)  $\lambda_1=0.03\cdot\Delta E_0/\omega$ is used. (c) $\lambda_1=0.025\cdot\Delta E_0/\omega$ is used. (d) Scan over various $\lambda_1$. \label{fig:res}}
\end{figure}

The shift of the resonances due to interactions is demonstrated using MCTDHB calculations (solid lines in Figs.~\ref{fig:res} (a-c)), where $\la J_z\ra_T$ is shown for several interaction strengths $U_0 N=1$, $U_0 N=2$, and $U_0 N=4$, respectively. 
With increasing interaction strength, the resonances become shifted towards each other. For relatively strong interactions $U_0 N=4$ (figure~\ref{fig:res} (c)), the initial state is not completely localized and the contrast of the resonances is reduced.

The dependence of the resonance shift on the amplitude $\lambda_1$ is shown in figure~\ref{fig:res} (d) for $U_0 N=1$, calculated from the TM model. The shift  becomes smaller for larger $\lambda_1$. The black line corresponds to $\lambda_1=0.035\cdot \Delta E_0/\omega$, which was used in Figs.~\ref{fig:res} (a-c) to map out the resonance most clearly.

For $U_0 N=1$ and for the TM model, we compare also to the case of artificially driving only with $\Omega(t)$ or $\Delta E(t)$, similar as in figure~\ref{fig:enhanced}. We find that the \emph{enhanced Shapiro resonances} are much more distinct than the usual Shapiro resonances, and the difference is much larger than in the non-interacting case. We also find that the resonance shifts of the usual Shapiro resonances (i.e., at constant $\Omega$) are different from those of the enhanced Shapiro resonances and the case of constant $\Delta E$. This is similar to the findings in~\cite{zhang:08}, where it has been shown that an oscillating $\Omega(t)$ leads to a shift in the effective interaction strength.

\subsection{Spatial dynamics and damping \label{subsec:damp}}

Finally we discuss the differences between several models for calculating the resonances. The best description is provided by the MCTDHB method, such that the MCTDHB results in Figs.~\ref{fig:res} (a-c) (solid lines) serve as a reference. The main question we want to answer here are the limitations of simpler models in describing the driven system. 

We compare in Figs.~\ref{fig:res} the resonance structure for (a) $U_0 N=1$, (b) $U_0 N=2$, and (c) $U_0 N=4$ calculated within different models. For $U_0 N=1$ we find that there is a very good agreement between MCTDHB (solid line), GP (dashed-dotted line) and TM model (dashed line), a discrepancy only exists regarding the widths of the resonances. 
For $U_0 N=2$ the GP fails to be a good description, especially for the first resonance. For $U_0 N=4$ finally, the TM model does neither correctly reproduce depth and location of the resonances. The GP equation gives slightly better results than the TM-model, but still appreciably misses the correct results. From these simulations we see that, whereas for $U_0 N=1$ and $U_0 N=2$ the coincidence between TM and MCTDHB is very good for all times, for $U_0 N=4$ the coincidence is lost at short times. Thus, whereas for weaker interactions $\Omega(t)$ and $\Delta E(t)$ as deduced from static orbitals are sufficient, the full self-consistency of MCTDHB is needed for stronger interactions ($U_0 N=4$).

\begin{figure}
\centering
\begin{tabular}{l l}
\includegraphics[width=.45\columnwidth]{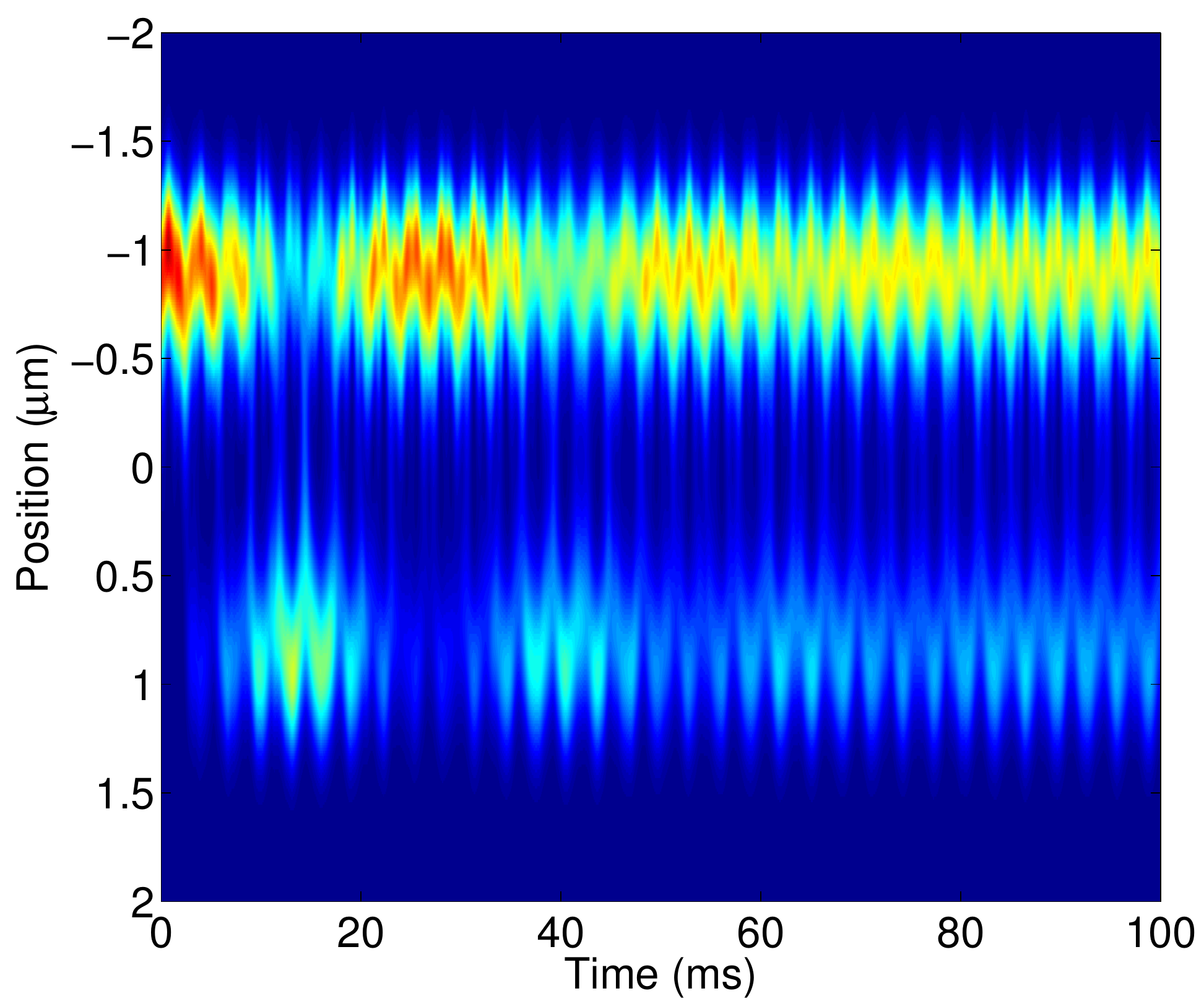}&\includegraphics[width=.45\columnwidth]{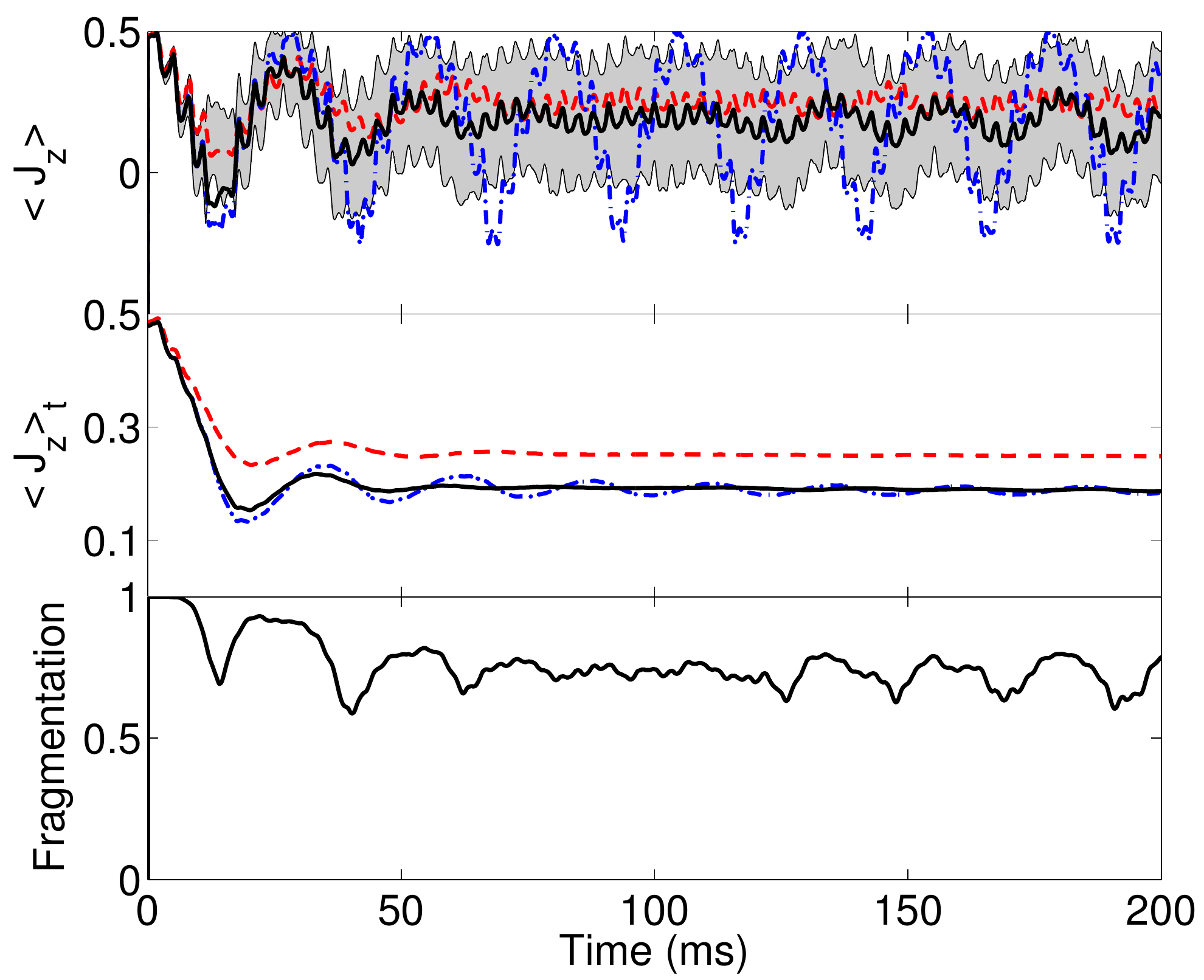}\\
(a)&(b)\\
 \includegraphics[width=.45\columnwidth]{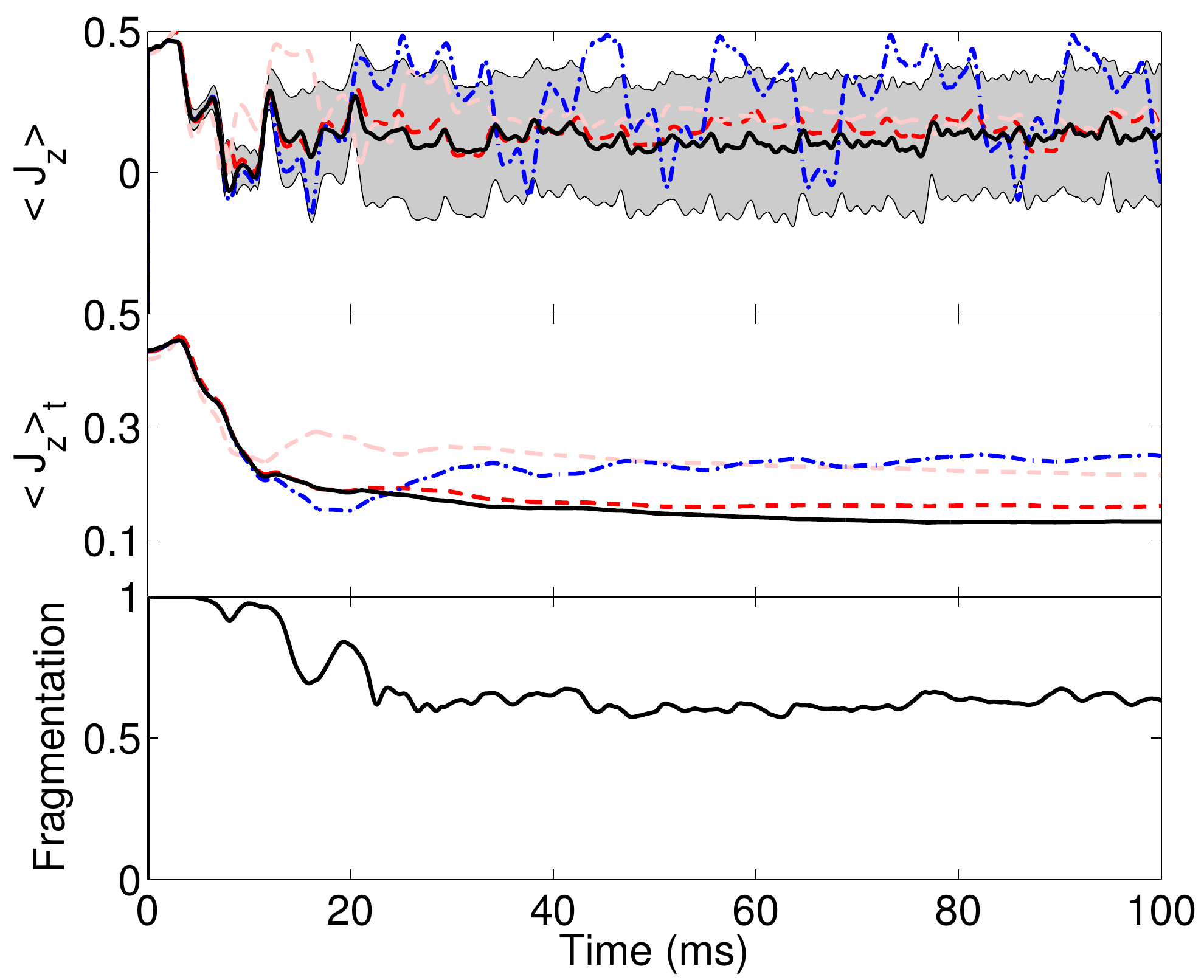}&\includegraphics[width=.45\columnwidth]{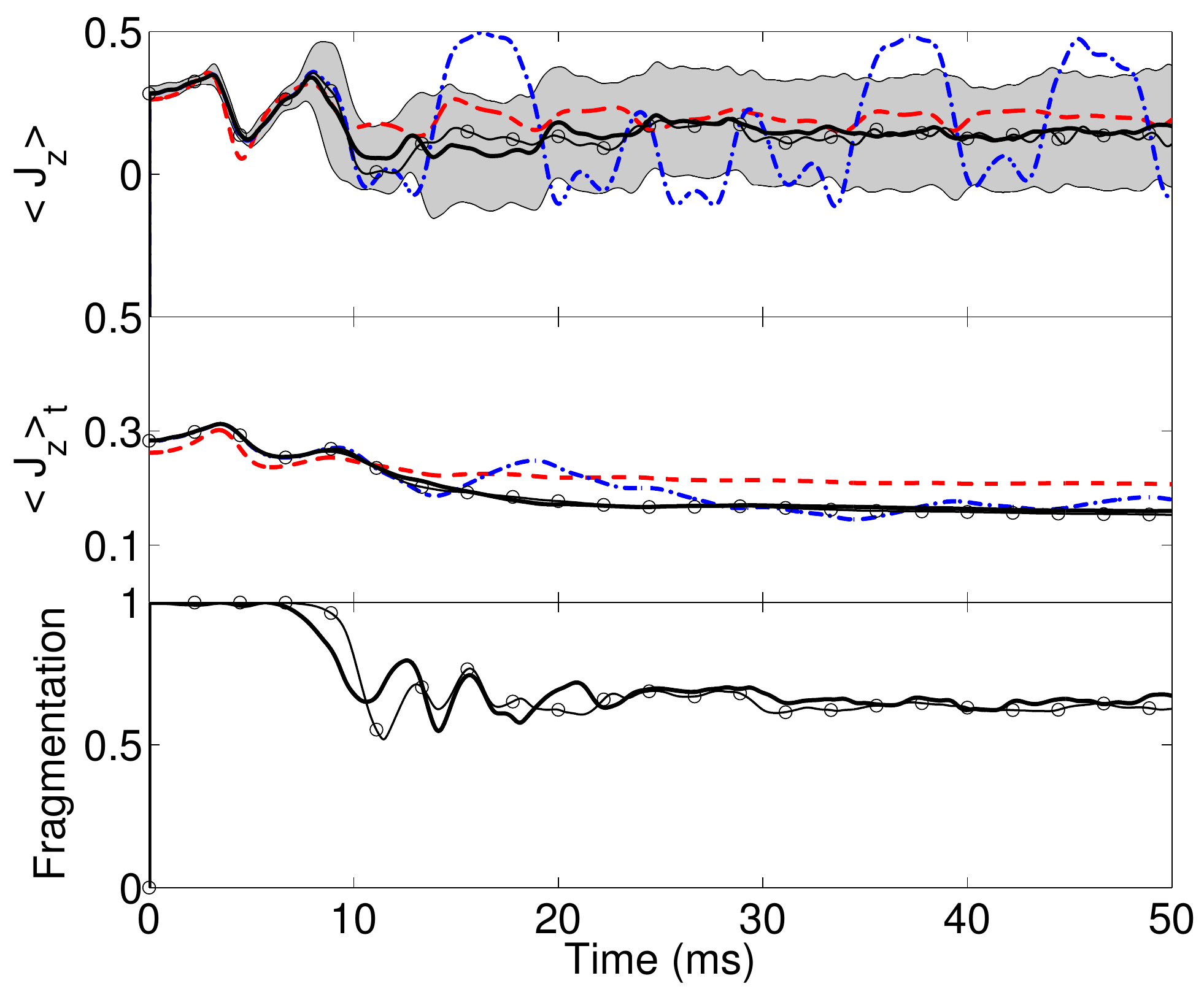}\\
 (c)&(d)
\end{tabular}
\caption{(a) Density for $U_0 N=1$ and driving frequency $\omega/\Delta E_0=0.85$.  ($N=100$) (b-d) Example trajectories for different interaction strengths ($N=100$), with color code as in figure~\ref{fig:res}. The upper panel represents $\la J_z\ra(t)$, where the gray band represents the variance $\Delta J_z$ for MCTDHB, the middle panel the averaged quantity $\la J_z\ra_T$, and the lower panel the fragmentation $(\rho_{11}-\rho_{22})/N$. (b) $U_0 N=1$ and $\omega/\Delta E_0=0.85$. (c) $U_0 N=2$ and $\omega/\Delta E_0=0.6$.  (c) $U_0 N=4$ and $\omega/\Delta E_0=0.52$. The gray line represents an MCTDHB solution for $N=1000$ (with $U_0 N$ constant). \label{fig:damp}}
\end{figure}

The spatial dynamics of the condensates as shown in figure~\ref{fig:damp} (a) is adiabatic to some extent, since the driving frequency is much smaller than the transverse trap frequency $\Delta E_0\ll\omega_{x,y}$.   
However, as we have seen, for strong interactions ($U_0 N=4$) the exact details of the condensate oscillations become important and have to be taken into account.

We show in Figs.~\ref{fig:damp} (b-c) example trajectories for the first resonances of (a) $U_0 N=1$, (b) $U_0 N=2$, and (c) $U_0 N=4$, with same color code as in figure~\ref{fig:res}. The upper panels show the atom number imbalance $\la J_z\ra(t)$, whereas the middle panels show the time averaged atom number imbalance $\la J_z\ra_t=\frac{1}{t}\int_0^t dt'\la \hat J_z(t')\ra$. The trajectories of the TM model and MCTDHB mostly deviate, in particular for $U_0 N=4$, where the trajectory of the TM model differs completely from the MCTDHB one after $10$\,ms. In order to point out the importance of the renormalization to the tunnel coupling according to the last term in equation~\ref{eq:improved}, we also indicate results where it is not included (bright dashed line figure~\ref{fig:damp} (c)).

The GP equation is in principle not a good description of the Shapiro dynamics, as it does not take into account damping of the Shapiro oscillations on a time scale of tens of ms.  We show in the upper panels of Figs.~\ref{fig:damp} (b-c) the width of the number distribution $\Delta J_z(t)$ for the TM calculations by the bright band around the mean value $\la J_z\ra(t)$. Whereas for short times the number distribution is very narrow, it starts to become much broader after a time which depends on interactions. At the same time the GP results start to deviate from the exact ones. Finally, the distribution stays very broad and the mean value does not really change anymore, while the GP equation predicts oscillations of $\la J_z\ra(t)$ with approximately constant amplitude. 

Thus, the time averaged imbalance $\la J_z\ra_T$ does not fully characterize the system dynamics, and it is more appropriate to study the time evolution of the imbalance. However, it has been expressed in the literature~\cite{eckardt:05} that for weak interactions and the "standard" Shapiro effect, the averaged imbalance $\la J_z\ra_T$ of the TM model and the GP equation should be the same, although the dynamics are very different in each case. For the configuration we discuss here we find this to be justified only for $U_0 N=1$ (see figure~\ref{fig:res} (a)). In this case the damping is slower than the period of the oscillations, and $\la J_z\ra_T$ corresponds approximately to the average of the extremal values, see the middle panel of figure~\ref{fig:damp} (b). However, for stronger interactions, the imbalance becomes stuck at an early point of the dynamics and thus the results for the averaged imbalance are significantly different from those predicted by the GP equation. 

In order to have another perspective on the dynamics of damping, we plot the population difference of the natural orbitals $\phi_i(x)$ ($i=1,2$).  As can be seen in the lower panels of figure~\ref{fig:damp}, during the Shapiro dynamics the BECs become fragmented into two incoherent ($\la \hat a_1^{\dagger} \hat a_2\ra=0$) condensates due to the nonlinear interactions. We find that indeed the GP description starts to fail whenever the system starts to fragment. 

In addition we plot in figure~\ref{fig:damp} (d) also results for $N=1000$ atoms (thin line with symbols), in order to demonstrate that fragmentation is clearly important also for larger numbers of atoms. Thus, convergence of MCTDHB results to GP results happens only at very large $N$. \footnote{The interaction strength considered in this work are typically realized with $N=100$ to $N=1000$ atoms.}

\section*{Conclusion}

In conclusion, we have studied Shapiro resonances in a configuration where not only the bias potential, but also the tunnel coupling is driven dynamically. This is typical for double well potentials realized  on atomchips and thus our findings are directly relevant for future experiments. We show that this configurations has favorable properties as it leads to  \emph{enhanced Shapiro resonances}. Due to a spatial deformation of the potential induced by the driving, the question of transverse excitations is of great interest. We find that, at least for significant interactions, the realistic MCTDHB method has to be used instead of simpler models in order to properly capture spatial dynamics of the involved modes.

\section*{Acknowledgements}
We thank Alexej I. Streltsov, Ofir E. Alon, Lorenz S. Cederbaum, Aurelien Perrin, and Tarik Berrada for most helpful discussions. Special thanks go to Axel U. J. Lode for explanations and help regarding the MCTDHB package. This work has been supported in part by NAWI GASS and the ESF Euroscores programm: EuroQuaser project QuDeGPM. We also acknowledge support by the Austrian Science Fund within projects P21080-N16, P22590-N16 and F41, by the EU project MIDAS, and by the Viennese Fund for Science and Technology (WWTF) projects MA45 and MA07-7. J.G. acknowledges support by the Humboldt-foundation.

\end{document}